\documentclass[10pt,letterpaper,twocolumn]{article}

\usepackage{ol2}
\usepackage[draft]{hyperref}
\usepackage{amsmath}

\begin{document}

\twocolumn[

\title{Nonreciprocal switching thresholds in coupled nonlinear microcavities}

\author{Victor Grigoriev$^*$ and Fabio Biancalana}

\address{
Max Planck Institute for the Science of Light, G\"{u}nther-Scharowsky-Str. 1, Bau 26, Erlangen 91058, Germany
$^*$Corresponding author: victor.grigoriev@mpl.mpg.de
}

\renewenvironment{abstract}%%
{
\noindent \begin{center} 
{\footnotesize Compiled \today}
\vskip4pt \begin{minipage}{34.25pc} \parindent.2in \noindent \footnotesize \rm
}
{
\\ \\ \hfil \end{minipage} \end{center}
}   

\begin{abstract}
A novel concept for the design of nonlinear optical diodes is proposed which uses the multistability of coupled nonlinear microcavities and the dependence of switching thresholds on the direction of incidence. A typical example of such diode can be created by combining two mirror symmetric microcavities where modes of the opposite parity dominate. It is shown that a strong nonreciprocal behavior can be achieved together with a negligible insertion loss. To describe the dynamical properties of such systems, a model based on the coupled mode theory is developed, and a possible implementation in the form of multilayered structures is considered.
\end{abstract}


]

%\section{Introduction}
\noindent
The development of integrated photonic circuits requires all-optical elements for high-speed processing of light signals. The optical diode (OD) is one of such indispensable elements. Similar to electronic diodes, it allows the flow of light only in one direction and can reduce problems caused by unwanted reflections or interference effects. The successful design of an OD relies on the breaking of time-reversal symmetry \cite{Potton2004}. The Faraday effect in magneto-optical media is mostly used for this purpose. There has been a considerable progress during recent years in making such devices suitable for integration into waveguides on a chip \cite{Doetsch2005, Takeda2008, Jalas2010}. However, the need to apply a strong external magnetic field still remains a major limiting factor for their development. Alternative approaches to fabricate ODs are based on chiral media such as cholesteric liquid crystals \cite{Hwang2005} or metamaterials \cite{Menzel2010, Zhukovsky2009}. Although their figures of merit are significantly lower, they can be produced with very small sizes and do not require any external fields.

Nonlinear effects attracted a lot of attention as a means of achieving unidirectional transmission. They can be subdivided into two large groups. The first one uses the nonreciprocal conversion of waveguide modes due to asymmetrically placed defects in directional couplers \cite{Alberucci2008} and quasi-phase-matched gratings \cite{Gallo2001, Yu2010}. As an option, the asymmetry can be created by launching additional signals that modulate the refractive index dynamically, leading to indirect photonic transitions \cite{Yu2009}. The second group uses the folding of resonances in microcavities (MCs) possessing the Kerr nonlinearity and can be implemented in many different configurations \cite{Chremmos2010}: from photonic crystals with embedded defects of various form and dimensionality \cite{Soljacic2003, Zhao2006, Cai2008} to quasiperiodic \cite{Biancalana2008, Grigoriev2010} and aperiodic \cite{Shadrivov2010, Zhukovsky2011} structures.

The resonant properties of the MCs allow ODs of smaller sizes to be constructed while keeping a large contrast ratio between the transmission in the forward and backward directions. However, a common problem in the designs based on the usage of MCs is that the high contrast ratio is often achieved at the expense of lowering the maximal transmission in the forward direction. In this Letter, we show how to solve this problem and to create nonlinear ODs with a negligible insertion loss.

\begin{figure}[b]
\centerline{\includegraphics[width=80mm]{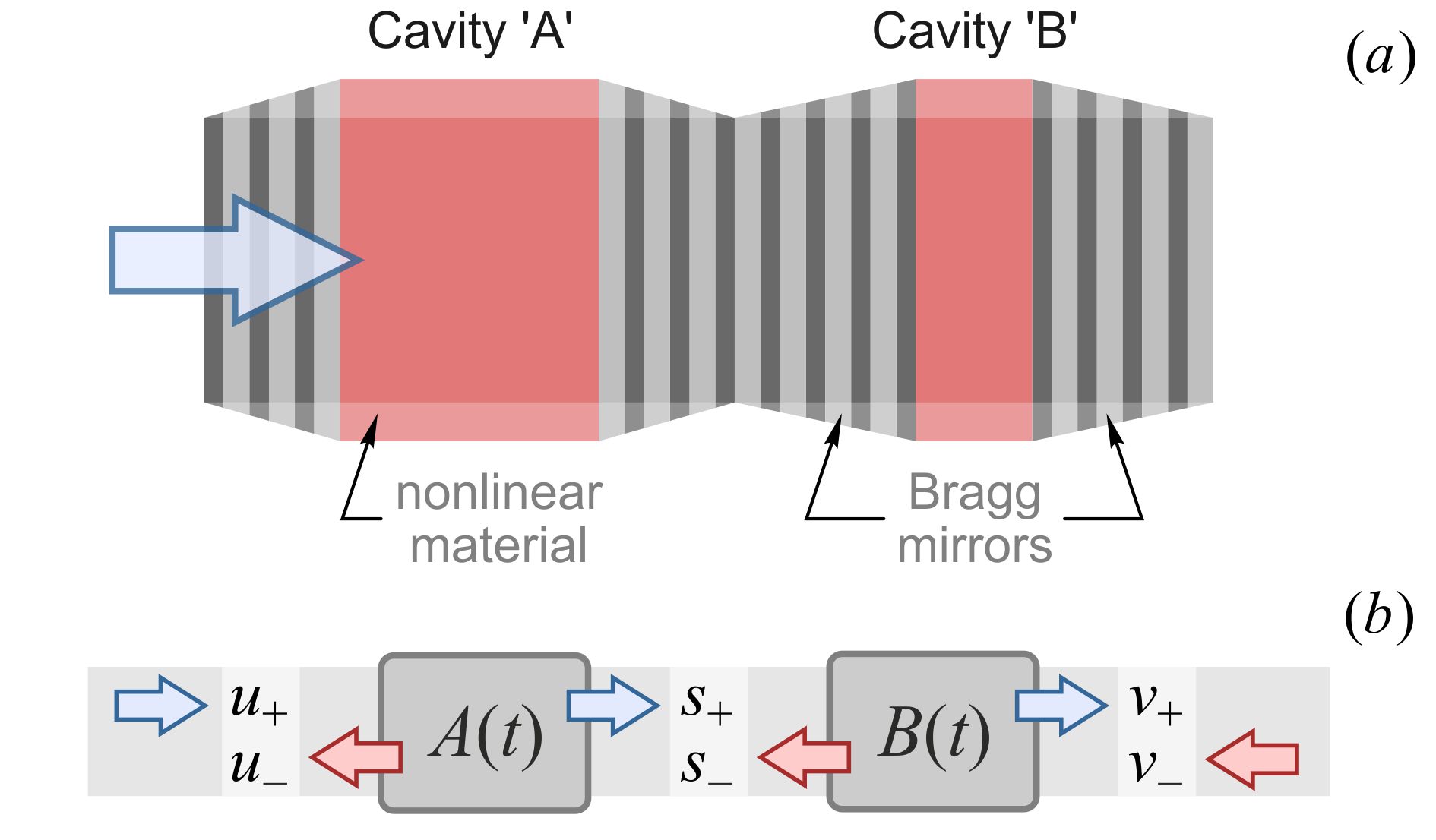}}
\caption{\small{\label{fig1} (Color online)
(a)~A schematic representation of the proposed OD in the form of multilayered structure. It consists of two MCs filled with nonlinear material.
(b)~A sketch of the model used to describe the transient dynamics.
}}
\end{figure}

%\section{Model}
It is known that the mirror symmetric MCs demonstrate resonances of perfect transmission. They can be used as a building block to construct more complex structures, and if these MCs are designed to have a common resonance of perfect transmission, the resulting structure will also show it in the linear spectrum \cite{Zhukovsky2010}. To create asymmetry, it is sufficient to use two such MCs and to modify their sensitivity to the intensity of incident waves. Without any loss of generality, we can consider a multilayered geometry shown in Fig.~\ref{fig1}(a). In this case, the nonlinear response of each MC can be changed either directly by varying the thickness of the nonlinear defect region or by modifying the localization strength via the thickness of the Bragg mirrors. These two degrees of freedom can be rigorously taken into account in the framework of coupled-mode theory \cite{Bravo-Abad2007, Haus1984}, and we will investigate how to use them efficiently to preserve the perfect transmission in the nonlinear case.

The coupled-mode equations for the mirror symmetric MC can be written as
\begin{equation}
\label{eqA}
\frac {dA} {dt} =
    - \left[ i
            \left(
                \omega_0 - \gamma_{ \rm{A} } \frac {|A|^2} {I_{ \rm{A} }}
            \right)
            + \gamma_{ \rm{A} }
      \right] A
    + \gamma_{ \rm{A} } (u_+ \pm s_- ),
\end{equation}
\begin{equation}
\label{eqUV}
\left( \begin{array}{c}
    u_- \\ s_+
\end{array} \right)
 = \mp
\left( \begin{array}{c}
    u_+ \\ s_-
\end{array} \right)
 + A
\left( \begin{array}{c}
 \pm1 \\ 1
\end{array} \right),
\end{equation}
where $A(t)$ is the amplitude of the mode $E_{ \rm{A} }(x)$ with the resonant frequency $\omega_0$ and damping constant $\gamma_{ \rm{A} }$, $u_\pm$($s_\pm$) are the amplitudes of the forward and backward propagating plane waves on the left (right) side of the MC [Fig.~\ref{fig1}(b)]. The Kerr nonlinearity enters via the characteristic intensity $I_{ \rm{A} }$ and is responsible for the nonlinear shift of resonant frequencies. If the mode $E_{ \rm{A} }(x)$ is even (odd), the lower (upper) signs should be used in these equations.

It is convenient to go into the frequency domain and to derive the transfer matrix ${\bf{M}}_{\rm{A}}$ which relates the plane waves on the opposite sides of the MC as $[u_+, \; u_- ]^{\rm{T}} = {\bf{M}}_{\rm{A}} [s_+, \; s_- ]^{\rm{T}}$  where
\begin{equation}
\label{eqTmatrix}
{\bf{M}}_{\rm{A}} =
{\bf{I}} - i
\left(
 \frac {\omega - \omega_0} {\gamma_{ \rm{A} }}
 + \frac {
    \left|
         s_+ \pm s_-
    \right|
    ^2 } {I_{ \rm{A} }}
\right)
\left[
    {\begin{array}{cc}
       1 &  \pm 1  \\
       \mp 1 &  -1  \\
    \end{array}}
\right]
\end{equation}
%    %
%\begin{equation}
%\label{eqTmatrixDef}
%\left( \begin{array}{c}
%   u_+ \\ u_-
%\end{array} \right) =
%{\bf{M}}_{\rm{A}}
%\left( \begin{array}{c}
%   s_+ \\ s_-
%\end{array} \right),
%\end{equation}
%    %
and $\bf{I}$ is the identity matrix. Regardless of the nonlinear term, the form of the transfer matrix (\ref{eqTmatrix}) satisfies such fundamental physical properties as the time-reversal symmetry (${\bf{M}}_{22} = {\bf{M}}_{11}^*$, ${\bf{M}}_{12} = {\bf{M}}_{21}^*  $), the conservation of energy flow ($\det {\bf{M}} = 1$) and in addition the spatial mirror symmetry (${\bf{M}}_{12} = - {\bf{M}}_{21}$).

The transmission spectrum for waves incident from the left can be found as $T_{\rm{A}} = |s_+ / u_+ |^2 = |({\bf{M}}_{\rm{A}})_{11}|^{-2}$
\begin{equation}
\label{eqTransmission}
T_{ \rm{A} } = \frac {I_{ \rm{out} }} {I_{ \rm{in} }} =
\left[
 1 + \left(
  \frac {\omega  - \omega_0} {\gamma_{ \rm{A} }} + \frac {I_{ \rm{out} }} {I_{ \rm{A} }}
  \right) ^ 2
\right] ^ {-1}.
\end{equation}
It follows from the formula (\ref{eqTransmission}) that the condition of perfect transmission can be written as $I_{\rm{in}} / I_{\rm{A}} = - \Delta \omega / \gamma_{\rm{A}}$, where $\Delta \omega = \omega - \omega_0$ is the frequency detuning. This condition can be generalized to the case of two or more MCs which have the same resonant frequency but differ in their characteristic intensities and quality factors $Q = \omega_0 / (2 \gamma )$
\begin{equation}
\label{eqCondition}
Q_{\rm{A}} I_{\rm{A}} = Q_{\rm{B}} I_{\rm{B}}.
\end{equation}
It is useful to introduce the parameter $f > 1$ as a measure of asymmetry between these MCs $f = Q_{\rm{A}} / Q_{\rm{B}} = I_{\rm{B}} / I_{\rm{A}}$. Since larger quality factors generally imply smaller characteristic intensities, the condition (\ref{eqCondition}) is not difficult to fulfill.

\begin{figure}[b]
\centerline{\includegraphics[width=80mm]{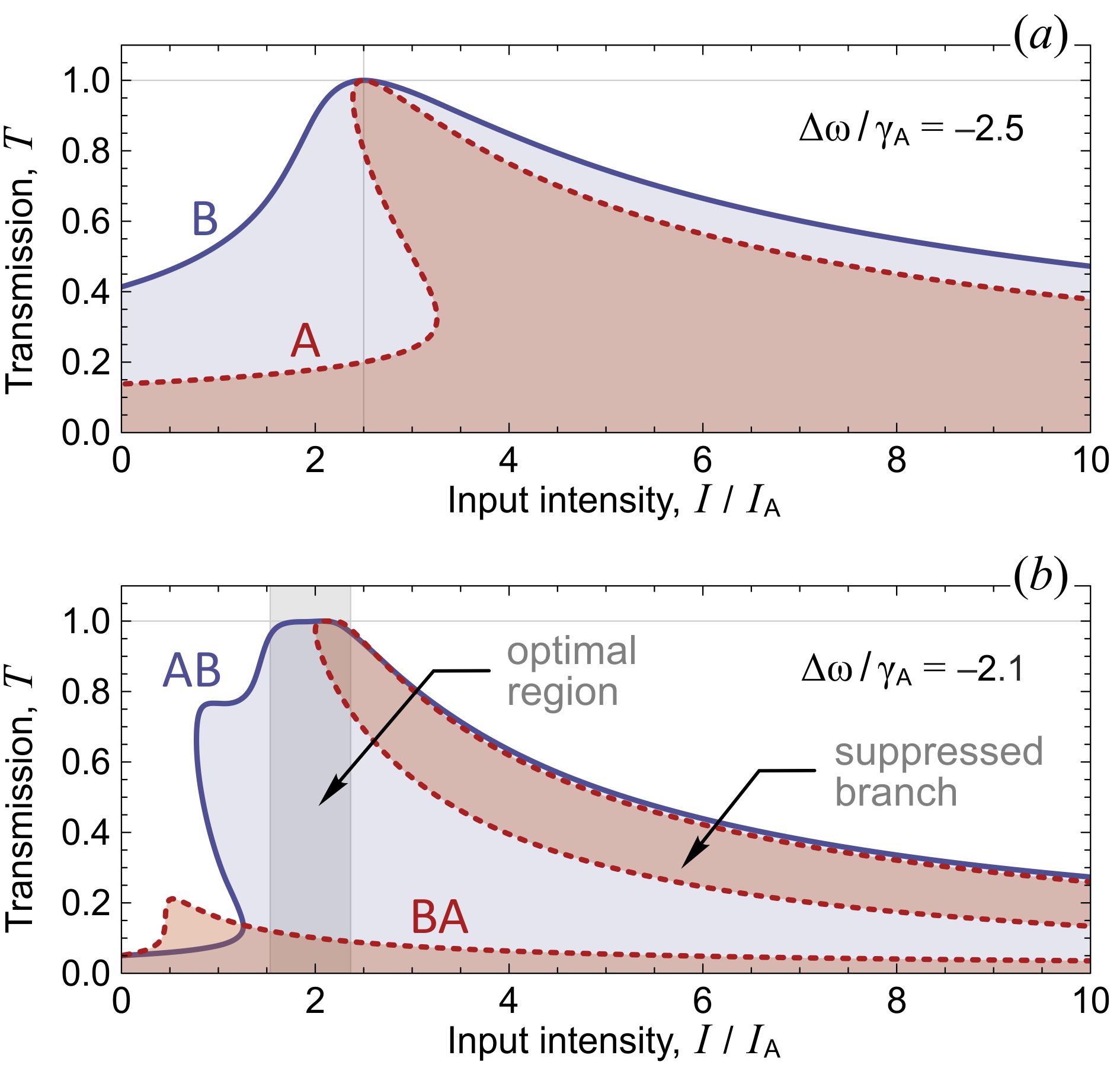}}
\caption{\small{\label{fig2} (Color online)
(a) The hysteresis of transmission for single MCs 'A' and 'B'. Their quality factors $ Q_{ \rm{B} } = Q_{ \rm{A} } / f $ and characteristic intensities $ I_{ \rm{B} } = f I_{ \rm{A} } $  are related by the parameter $ f = 2.1 $.  (b) The hysteresis of transmission for coupled MCs 'A' and 'B' when signals are propagating in the 'AB' and 'BA' directions.
}}
\end{figure}

%\section{Hysteresis}
The hysteresis of transmission for single MCs 'A' and 'B' can be computed by using Eq.~(\ref{eqTransmission}) and is shown in Fig.~\ref{fig2}(a). For any value of the parameter $f$ and the frequency detuning $\Delta \omega$, the resonances of perfect transmission are equally shifted by the Kerr nonlinearity and coincide in both MCs. However, the switching thresholds can be different, and a specific range of frequency detunings exists when only one of the MCs shows the bistable behavior.

\begin{figure}[b]
\centerline{\includegraphics[width=80mm]{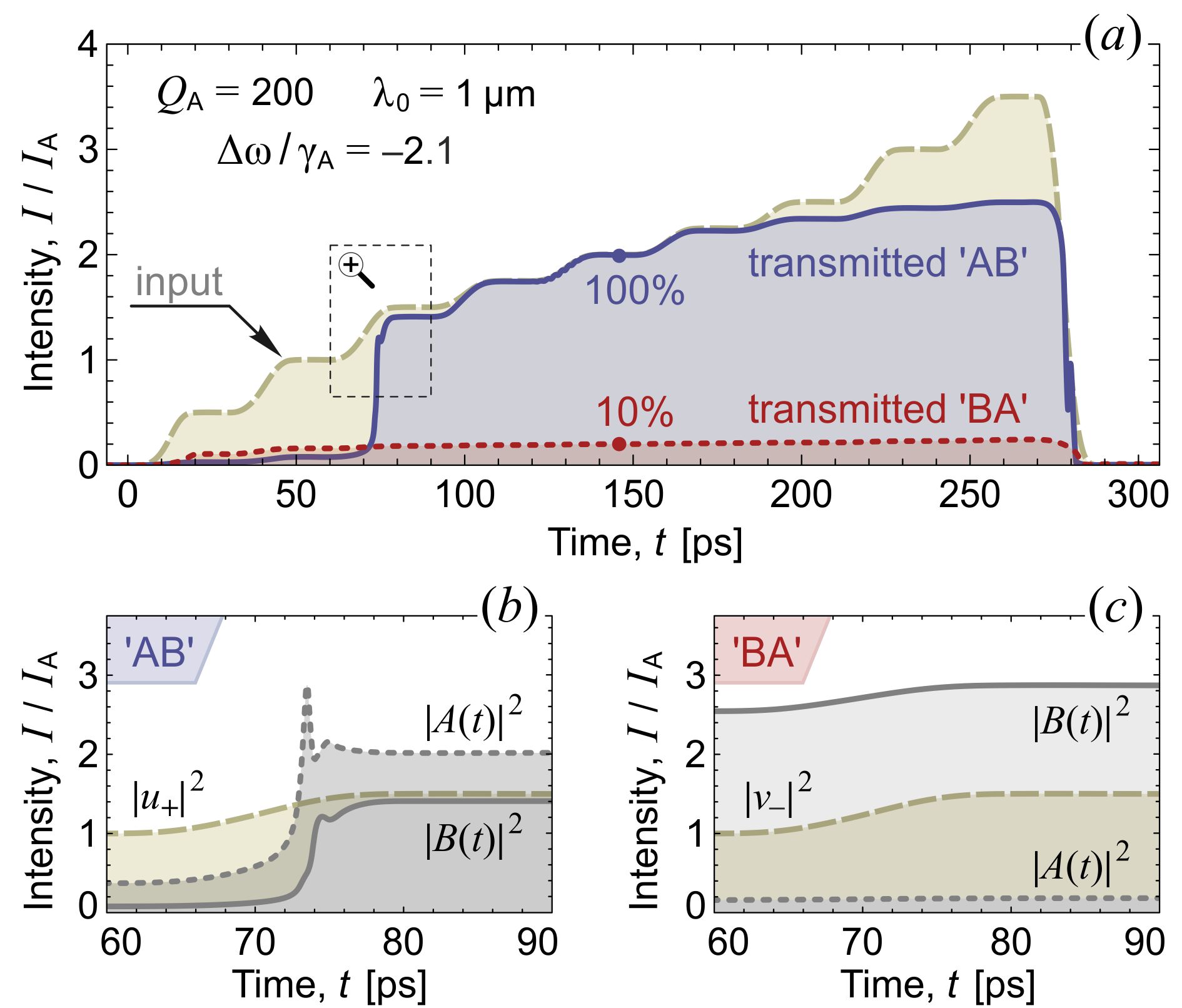}}
\caption{\small{\label{fig3} (Color online)
(a) The transient dynamics of the OD for a continuous wave signal with the carrier frequency $ \omega = \omega_0 - 2.1 \gamma_{ \rm{A} } $ and a staircase envelope. The intensities of the input and transmitted signals for the opposite directions of incidence are shown as a function of time.  (b) The amplitudes of the MCs during the switching to a high transmission state.  (c) When the input intensity is increased for the reversed direction of incidence, the amplitudes of the MCs move away from the resonance condition $|A(t)|^2 / I_{\rm{A}} = |B(t)|^2 / I_{\rm{A}} = - \Delta \omega / \gamma_{\rm{A}}$.
}}
\end{figure}

The hysteresis of transmission for coupled MCs 'A' and 'B' can be found similar to the linear transfer matrix method. After choosing some output intensity, the nonlinear transfer matrices $\bf{M}_{\rm{A}}$ and $\bf{M}_{\rm{B}}$ defined according to
Eq.~(\ref{eqTmatrix})
%Eqs.~(\ref{eqTmatrix})--(\ref{eqTmatrixDef})
can be computed in a reversed order, giving the total transfer matrix of the structure and the input intensity as a result. It turns out however that the matrices do not commute in the nonlinear case, and their multiplication leads to a different answer depending on the direction of incidence [Fig.~\ref{fig2}(b)]. The effect is maximized when the MCs are of the opposite parity and can be explained in the following way. After reaching the second MC in the system, a part of the signal is reflected and increases the field in the first MC due to constructive interference. When 'A' is used as the first MC, the small increase causes the switching inside it, and the transmission through both MCs is greatly enhanced. On the contrary, when 'B' is used as the first MC, the small increase draws it away from the resonance so that the switching to a higher transmission state is strongly suppressed. To confirm these considerations, we performed a time domain simulation [Fig.~\ref{fig3}].

In comparison to other approaches based on the usage of MCs, the proposed design of OD offers several important advantages. First of all, it shows not only a low insertion loss, but also it is stable for small variations of the input intensity and frequency detuning. This is mostly because the resonance of perfect transmission changes its shape and flattens around the optimal set of parameters. Secondly, the switching thresholds for the opposite propagation directions are separated considerably. This prevents the excitation of a higher transmission branch on the hysteresis curve by waves moving in the blocked direction. At the same time, the waves moving in the allowed direction can cause an immediate switching to a perfect transmission even if they have relatively small intensities.

We found that the above properties are more pronounced when the parameter $f$ is in the following range $2 < f < 3$ and the typical contrast ratio between the transmission in the opposite directions varies from $10$ to $20$. Although this number is not very large, the low insertion loss makes a series connection of such diodes to be very favorable so that arbitrary high transmission contrasts can be achieved in principle.

%\section{Example}
The OD can be implemented in the form of a multilayered structure [cf.~Fig.~\ref{fig1}(a)]. Using layers of the quarter-wave optical thickness, the corresponding symbolic formula can be written as
\begin{equation}
\label{eqSymbolic}
%\underbrace {
    ({ \rm{HL} }) ^ {p_{ \rm{A} }}
    { \rm{D} }    ^ {2m_{ \rm{A} }}
    ({ \rm{LH} }) ^ {p_{ \rm{A} }}
%}_{ \rm{Cavity \; A} }
\;
%\underbrace {
    ({ \rm{LH} }) ^ {p_{ \rm{B} }}
    { \rm{D} }    ^ {2m_{ \rm{B} }}
    ({ \rm{HL} }) ^ {p_{ \rm{B} }}
,%}_{ \rm{Cavity \; B} },
\end{equation}
where 'H' ('L') denotes layers with higher (lower) linear refractive index, 'D' is a defect layer made from a Kerr-nonlinear material, $p_{\rm{A}(\rm{B})}$ and $m_{\rm{A}(\rm{B})}$ are integers. The wavelength used for the quarter wave condition determines the resonance in the vicinity of which the nonreciprocal behavior can be observed. For the optimal performance, the linear refractive indices should satisfy the following relation $(n_{\rm{H}} / n_{\rm{L}}) ^ {p_{\rm{B}}- p_{\rm{A}}} = n_{\rm{D}}$. In this case, the characteristic intensities will be inversely proportional to the width of defect regions, and the parameter $f$ can be found as $f = m_{\rm{A}} / m_{\rm{B}}$.

%\section{Conclusions}
In conclusion, we proposed how to create an efficient OD by combining two nonlinear MCs of opposite parity. Our design shows a strong nonreciprocal behavior together with a negligible insertion loss and can be fabricated in a variety of configurations such as multilayered structures or photonic crystals with embedded defects.

~%Acknowledgements

This work was supported by the German Max Planck Society for the Advancement of Science
(MPG).

%\bibliography{references}

\begin{thebibliography}{10}
\newcommand{\enquote}[1]{``#1''}

\bibitem{Potton2004}
R.~J. Potton, Rep. Progr. Phys. \textbf{67}, 717 (2004).

\bibitem{Doetsch2005}
H.~D\"{o}tsch, N.~Bahlmann, O.~Zhuromskyy, M.~Hammer, L.~Wilkens, R.~Gerhardt,
  P.~Hertel, and A.~F. Popkov, J. Opt. Soc. Am. B \textbf{22}, 240 (2005).

\bibitem{Takeda2008}
H.~Takeda and S.~John, Phys. Rev. A \textbf{78}, 023804 (2008).

\bibitem{Jalas2010}
D.~Jalas, A.~Petrov, M.~Krause, J.~Hampe, and M.~Eich, Opt. Lett. \textbf{35},
  3438 (2010).

\bibitem{Hwang2005}
J.~Hwang, M.~H. Song, B.~Park, S.~Nishimura, T.~Toyooka, J.~W. Wu,
  Y.~Takanishi, K.~Ishikawa, and H.~Takezoe, Nat. Mater. \textbf{4}, 383
  (2005).

\bibitem{Menzel2010}
C.~Menzel, C.~Helgert, C.~Rockstuhl, E.~B. Kley, A.~Tunnermann, T.~Pertsch, and
  F.~Lederer, Phys. Rev. Lett. \textbf{104}, 253902 (2010).

\bibitem{Zhukovsky2009}
S.~V. Zhukovsky, A.~V. Novitsky, and V.~M. Galynsky, Opt. Lett. \textbf{34},
  1988 (2009).

\bibitem{Alberucci2008}
A.~Alberucci and G.~Assanto, Opt. Lett. \textbf{33}, 1641 (2008).

\bibitem{Gallo2001}
K.~Gallo, G.~Assanto, K.~R. Parameswaran, and M.~M. Fejer, Appl. Phys. Lett.
  \textbf{79}, 314 (2001).

\bibitem{Yu2010}
Z.~Y. Yu, F.~Xu, X.~W. Lin, X.~S. Song, X.~S. Qian, Q.~Wang, and Y.~Q. Lu, Opt.
  Lett. \textbf{35}, 3327 (2010).

\bibitem{Yu2009}
Z.~F. Yu and S.~H. Fan, Nat. Photonics \textbf{3}, 91 (2009).

\bibitem{Chremmos2010}
I.~Chremmos, O.~Schwelb, and N.~Uzunoglu, \emph{Photonic Microresonator
  Research and Applications}, Springer Series in Optical Sciences (Springer,
  New York, 2010).

\bibitem{Soljacic2003}
M.~Solja\v{c}i\'{c}, C.~Luo, J.~D. Joannopoulos, and S.~H. Fan, Opt. Lett.
  \textbf{28}, 637 (2003).

\bibitem{Zhao2006}
N.~S. Zhao, H.~Zhou, Q.~Guo, W.~Hu, X.~B. Yang, S.~Lan, and X.~S. Lin, J. Opt.
  Soc. Am. B \textbf{23}, 2434 (2006).

\bibitem{Cai2008}
X.~H. Cai, X.~S. Lin, and S.~Lan, Chin. Phys. Lett. \textbf{25}, 2085 (2008).

\bibitem{Biancalana2008}
F.~Biancalana, J. Appl. Phys. \textbf{104}, 093113 (2008).

\bibitem{Grigoriev2010}
V.~Grigoriev and F.~Biancalana, New J. Phys. \textbf{12}, 053041 (2010).

\bibitem{Shadrivov2010}
I.~V. Shadrivov, K.~Y. Bliokh, Y.~P. Bliokh, V.~Freilikher, and Y.~S. Kivshar,
  Phys. Rev. Lett. \textbf{104}, 123902 (2010).

\bibitem{Zhukovsky2011}
S.~V. Zhukovsky and A.~G. Smirnov, Phys. Rev. A \textbf{83}, 023818 (2011).

\bibitem{Zhukovsky2010}
S.~V. Zhukovsky, Phys. Rev. A \textbf{81}, 053808 (2010).

\bibitem{Bravo-Abad2007}
J.~Bravo-Abad, S.~Fan, S.~G. Johnson, J.~D. Joannopoulos, and
  M.~Solja\v{c}i\'{c}, J. Lightwave Technol. \textbf{25}, 2539 (2007).

\bibitem{Haus1984}
H.~A. Haus, \emph{Waves and Fields in Optoelectronics} (Prentice--Hall, New
  Jersey, 1984).

\end{thebibliography}
%\bibliographystyle{ol}
%\bibliographystyle{osajnl}

\end{document}